\documentclass[10pt,twocolumn,letterpaper]{article} 
\pdfoutput=1
\usepackage{times}  
\usepackage{helvet}  
\usepackage{courier}  
\usepackage{graphicx} 
\usepackage{caption} 
\usepackage{amssymb}
\usepackage{amsmath}

\usepackage{color}
\usepackage{algorithm}
\usepackage{algorithmic}

\usepackage{newfloat}
\usepackage{listings}
\usepackage[colorlinks,linkcolor=blue]{hyperref}
\usepackage{authblk}

\usepackage{booktabs}
\usepackage{multirow}
\usepackage{apacite}
\usepackage[authoryear]{natbib}
\date{}

\begin{document}
	
\title{Noise-Resistant Multimodal Transformer for Emotion Recognition}
\author[1]{Yuanyuan Liu}
\author[1,2]{Haoyu Zhang\footnote{Corresponding author}}
\author[4]{Yibing Zhan}
\author[5]{Zijing Chen}
\author[1]{Guanghao Yin}
\author[1]{Lin Wei}
\author[3]{Zhe Chen}

\affil[1]{School of Computer Science, China University of Geosciences (Wuhan)}
\affil[2]{National Engineering Research Center for Geographic Information System, China University of Geosciences (Wuhan)}
\affil[1,2]{\textit{\{liuyy, zhanghaoyu, ygh2, linw\}@cug.edu.cn}}

\affil[3]{School of Computer Science, The University of Sydney}
\affil[3]{\textit{zhe.chen1@sydney.edu.au}}

\affil[4]{JD Explore Academy, JD.com}
\affil[4]{\textit{zhanyibing@jd.com}}

\affil[5]{Peter Faber Business School, Australian Catholic University}
\affil[5]{\textit{zijing.chen@acu.edu.au}}

\maketitle
\begin{abstract}
Multimodal emotion recognition identifies human emotions from various data modalities like video, text, and audio. However, we found that this task can be easily affected by noisy information that does not contain useful semantics and may occur at different locations of a multimodal input sequence. To this end, we present a novel paradigm that attempts to extract noise-resistant features in its pipeline and introduces a noise-aware learning scheme to effectively improve the robustness of multimodal emotion understanding against noisy information. Our new pipeline, namely Noise-Resistant Multimodal Transformer (NORM-TR),  mainly introduces a Noise-Resistant Generic Feature (NRGF) extractor and a multimodal fusion Transformer for the multimodal emotion recognition task. In particular, we make the NRGF extractor learn to provide a generic and disturbance-insensitive representation so that consistent and meaningful semantics can be obtained. Furthermore, we apply a multimodal fusion Transformer to incorporate Multimodal Features (MFs) of multimodal inputs (serving as the key and value) based on their relations to the NRGF (serving as the query). Therefore, the possible insensitive but useful information of NRGF could be complemented by MFs that contain more details, achieving more accurate emotion understanding while maintaining robustness against noises. To train the NORM-TR properly, our proposed noise-aware learning scheme complements normal emotion recognition losses by enhancing the learning against noises. Our learning scheme explicitly adds noises to either all the modalities or a specific modality at random locations of a multimodal input sequence. We correspondingly introduce two adversarial losses to encourage the NRGF extractor to learn to extract the NRGFs invariant to the added noises, thus facilitating the NORM-TR to achieve more favorable multimodal emotion recognition performance. In practice, extensive experiments can demonstrate the effectiveness of the NORM-TR and the noise-aware learning scheme for dealing with  both explicitly added noisy information and the normal multimodal sequence with implicit noises. On several popular multimodal datasets (\textit{e.g.,} MOSI, MOSEI, IEMOCAP, and RML), our NORM-TR achieves state-of-the-art performance and outperforms existing methods by a large margin, which demonstrates that the ability to resist noisy information in multimodal input is important for effective emotion recognition.  
\end{abstract}

\section{Introduction}
An accurate understanding of human emotions is beneficial for several applications, such as multimedia analysis, digital entertainment, health monitoring, human-computer interaction, \textit{etc}~\citep{DBLP:journals/ets/ShenWS07, DBLP:series/lncs/BealeP08, DBLP:journals/inffus/QianZMYP19, DBLP:journals/csur/DMelloK15}. Compared with traditional emotion recognition, which only uses a unimodal data source, multimodal emotion recognition that exploits and explores different data sources, such as visual, audio, and text, has shown 
significant advantages in improving the understanding of emotions \citep{DBLP:conf/emnlp/ZadehCPCM17,DBLP:conf/acl/TsaiBLKMS19,DBLP:conf/cvpr/LvCHDL21,DBLP:conf/mm/HazarikaZP20,DBLP:conf/mm/YuanLXY21}, including happiness, anger, disgust, fear, sadness, neutral, and surprise. 

Recently, most existing multimodal emotion recognition methods mainly focus on  multimodal data fusion, including tensor-based fusion methods~\citep{DBLP:conf/acl/MorencyLZLSL18, DBLP:conf/emnlp/ZadehCPCM17, DBLP:journals/corr/abs-2007-02038, DBLP:conf/mm/YuanLXY21} and attention-based fusion methods~\citep{DBLP:conf/aaai/ZhaoMGYXXHCK20, DBLP:conf/icassp/HuangTLLN20, DBLP:journals/taslp/ZhouDZWLL21}.  The tensor-based fusion methods aim to obtain a joint representation of data with different modalities via multilinear function calculation. For example, TFN~\citep{DBLP:conf/acl/MorencyLZLSL18} used Cartesian product operation to calculate the relationship between different modalities to obtain satisfactory performance. Since the computational complexity of the Cartesian product increases dramatically with the feature dimension and the number of modalities, its performance easily suffers from degradation if not using overwhelmingly large model capacities. LMF~\citep{DBLP:conf/emnlp/ZadehCPCM17} introduced low-rank matrix factorization operation to reduce the computational cost. However, LMF tends to reduce useful information, resulting in a decrease in model performance. The attention-based fusion methods mainly employ attention learning mechanisms to make multimodal information  interact with each other. For instance, \cite{DBLP:journals/taslp/ZhouDZWLL21} introduced attention learning to automatically calculate the importance weights of audio and video modalities so that obtaining effective emotion-related information. \cite{DBLP:conf/aaai/ZhaoMGYXXHCK20} proposed a new attention-based VAANET that integrated spatial, channel-wise, and temporal attentions for audio-video emotion recognition. Overall, most attentional methods tend to explore the importance of modalities for more effectively fusion, while ignoring the influence of useful and useless information inherent in the modality.

\begin{figure}[t]
	\centering
	\includegraphics[width=1.0\linewidth]{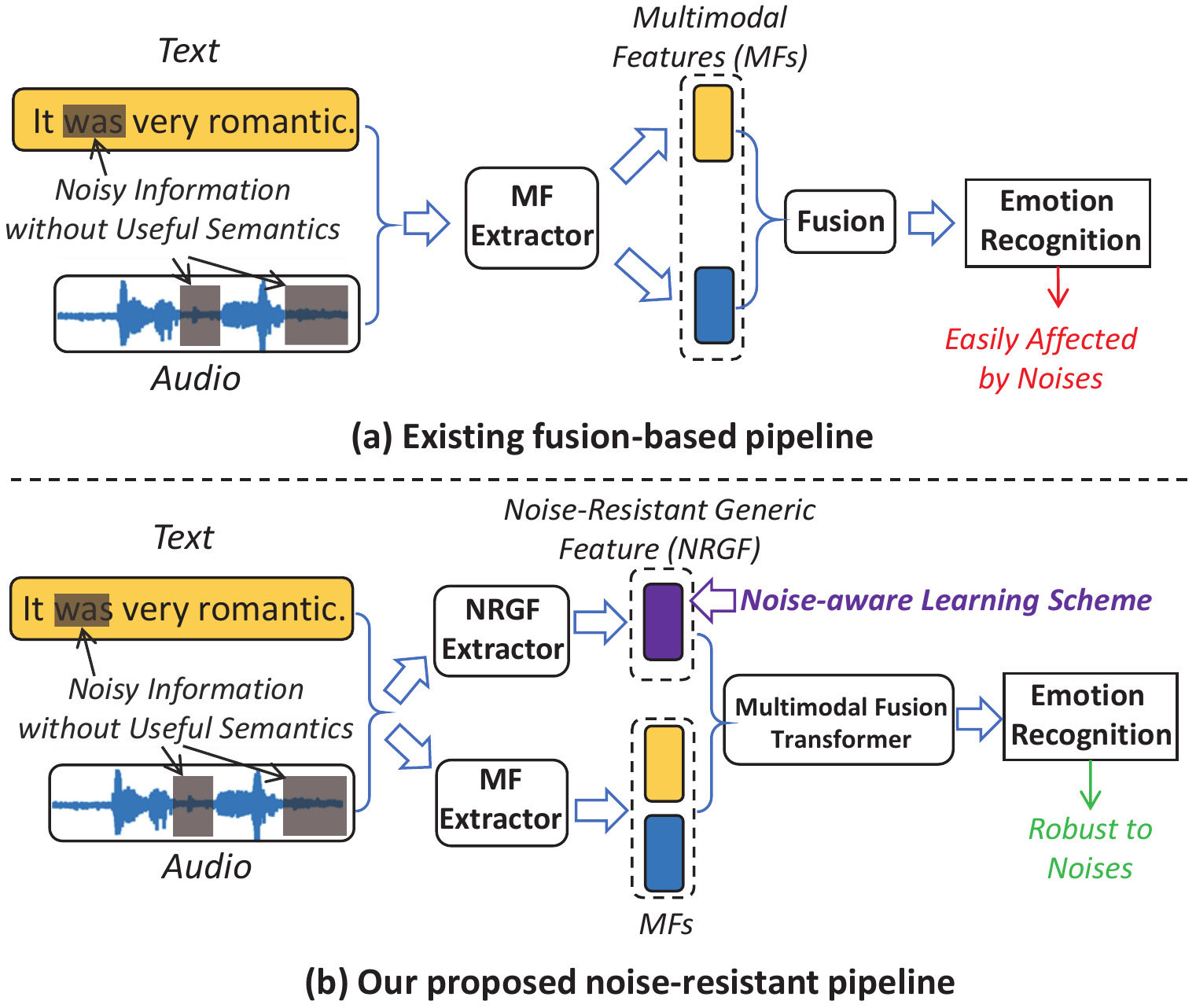}
	\caption{Our proposed multimodal emotion recognition methodology, \textit{i.e.}, Noise-Resistant Modality Transformer (NORM-TR) (as shown in \textbf{(b)}), compared to common multimodal fusion methodology (as shown in \textbf{(a)}). Using multimodal input that may contain noisy information with irrelevant useless semantics, existing multimodal emotion recognition method that directly fuses Multimodal Features (MFs) can be easily affected by the noises. Alternatively, we make our NORM-TR learn to extract Noise-Resistant Generic Feature (NRGF) with the help of a novel noise-aware learning scheme. Then, by using a multimodal fusion Transformer to make MFs complement the NRGF, we obtain much more robust multimodal emotion recognition results with our NORM-TR
	}
	\label{fig:motivation}
\end{figure}

Despite current progress in fusion strategies, we argue that alleviating the negative impacts of noisy information is also important. More specifically, we observe that, in a multimodal sequence, there could be plenty of information that shows little relevance to emotion understanding, which can be viewed as noisy information. For example, the background sounds in audio data are irrelevant to the human who smiles in the corresponding video. As a result, modeling the trivial information of these background sounds would likely affect the multimodal fusion and the final understanding performance. In our experiments, we can show that noisy information greatly degrades emotion recognition accuracy, which further implies that being insensitive to noises can be beneficial for accurate emotion understanding. However, to the best of our knowledge, current literature on multimodal emotion understanding lacks sufficient study on noisy information, thus still obtaining sub-optimal performance. 

In light of the above issue, we propose a novel Noise-Resistant Multimodal Transformer (NORM-TR) to address the adverse effects of noisy information on multimodal emotion recognition. The motivation of NORM-TR and the comparison to existing fusion methods are shown in Fig. \ref{fig:motivation}. In general, we make the NORM-TR learn to extract a Noise-Resistant Generic Feature (NRGF) and then apply a Transformer~\citep{DBLP:conf/nips/VaswaniSPUJGKP17} to incorporate Multimodal Features (MFs) extracted from the multimodal input according to their relations to the NRGF, thus obtaining more robust and more accurate emotion understanding results against noises. More specifically, we tend to formulate the NRGF to be generic and insensitive to the disturbances caused by noises. To obtain the NRGF, we employ an NRGF extractor and make it learn to summarize meaningful semantics from multimodal data. The extracted NRGF can provide a robust representation against noisy information, which also runs the risk of being insensitive to useful details for accurate emotion recognition. Therefore, we further introduce a multimodal fusion Transformer with NRGF serving as query and MFs serving as key and value; therefore, the relations between MFs and NRGF are reasoned, and the MFs can complement NRGF, achieving robust and accurate emotion recognition predictions.

To train the NORM-TR effectively, we first apply normal emotion recognition losses to make it learn to estimate human emotions. Meanwhile, we further apply our proposed noise-aware learning scheme to help make our model become robust against noises. Our noise-aware learning scheme explicitly adds noises to the input and applies adversarial losses to train the NRGF extractor in NORM-TR. Adding explicit noises can provide definite information about when noisy information occurs in the input, which can facilitate the discriminators of the related adversarial losses to be able to distinguish whether a feature contains noises. Specifically, two manners of adding noisy information are involved: 1) we make the added noisy information appear in all the modalities randomly, and 2) we add the noisy information to only a specific modality at some random periods. Two adversarial losses are introduced regarding both types of added noises, respectively. By fooling the discriminator that distinguishes the first type of added noises, we can make the NRGF extractor focus on more generic features against noises in all the multimodal data. Similarly, fooling the discriminator on the second type of added noises can make the extracted NRGF more robust against noises in each specific modality. Together with the emotion recognition loss, our proposed noise-aware learning scheme helps obtain robust NRGF and facilitates our NORM-TR model to achieve favorable multimodal emotion recognition performance. 

In summary, the major contributions of the paper can be described as: 
\begin{itemize} 
	\item We present a novel comprehensive study on noisy information for the multimodal emotion understanding task. To achieve more robust emotion understanding performance, we introduce the Noise-Resistant Multimodal Translator (NORM-TR) to extract Noise-Resistant Generic Features (NRGF) and significantly reduce the negative impacts of noise in the multimodal data. 
	\item Based on the NRGF, we devise a novel Transformer-based end-to-end pipeline for multimodal emotion recognition. Besides, a novel noise-aware learning scheme is further designed to help optimize the NORM-TR appropriately. 
	\item In practice, we demonstrate that the NORM-TR is effective in obtaining noise-invariant representations. Furthermore, our extensive experimental analysis of different popular datasets also illustrates that the NORM-TR significantly improves emotion recognition accuracy and achieves state-of-the-art performance by using the NRGF to alleviate the adverse impacts brought by noisy information, indicating the importance of handling noisy information. 
\end{itemize}

\section{Related Work}\label{sec:work}
\subsection{Multimodal Emotion Recognition}
Multimodal emotion recognition aims to predict human emotion from multiple modalities, such as video, audio, and text. Most existing methods \citep{DBLP:conf/mm/HazarikaZP20, DBLP:conf/aaai/ZhaoMGYXXHCK20, DBLP:conf/acl/TsaiBLKMS19, DBLP:conf/mm/YangHKDZ22, DBLP:conf/aaai/SunSSL20, DBLP:conf/iclr/TsaiLZMS19} mainly focus on how to learn and fuse multimodal emotion representations from data of different modalities by considering the difference and consistency of different modalities. For instance, \cite{DBLP:conf/mm/HazarikaZP20} proposed a multimodal representation learning method for modality-invariant and -specific subspace projection. Moreover, to learn the interactive information between different modalities, recent increasing work has concentrated on multimodal fusion mechanisms, where many elaborate multimodal fusion methods have been proposed. \cite{DBLP:conf/iclr/TsaiLZMS19} introduced Multimodal Factorization Mode (MFM) to explore intra-modal and cross-modal interactions by decomposing the modality representation into two independent sets of factors. \cite{DBLP:conf/aaai/SunSSL20} proposed Interaction Canonical Correlation Network (ICCN) that used Canonical Correlation Analysis (CCA)  to model the relationship between audio-text and video-text modalities. \cite{DBLP:conf/cvpr/LvCHDL21} proposed the Progressive Modality Reinforcement (PMR) approach to conduct multimodal fusion by considering the three-way interactions across all the involved modalities.

Although progress, current methods rarely focus on the noise problem in multimodal emotion recognition. \cite{DBLP:conf/mm/YuanLXY21} also pointed out that multimodal data contain a large amount of noise, such as missing data in some modal sequences, which can greatly degrade the results of multimodal fusion methods. How to effectively reduce the noise effect on multimodal data remains an open problem. 

\subsection{Adversarial Learning}
Adversarial learning is widely used in domain adaptation learning~\citep{DBLP:conf/icml/GaninL15, DBLP:conf/aaai/PeiCLW18, DBLP:journals/ijcv/WangHW20,he2022multi} and cross-modal retrieval~\citep{DBLP:conf/mm/WangYXHS17, DBLP:conf/cvpr/LiDL0GT18}, etc. Recently, to improve the effectiveness of fusion, adversarial learning is increasingly used in multimodal emotion recognition by learning common subspace representations. \cite{DBLP:conf/mm/YangHKDZ22} employed FDMER with adversarial learning to mine the commonality and diversity of different modalities, achieving an impressive performance for multimodal emotion recognition. Despite the progress, FDMER did not consider the side effect of noise on the robustness of the model. To address this limitation, we propose the NORM-TR with a novel noise-aware adversarial learning to extract Noise-Resistant Generic Features (NRGF), thereby greatly reducing the negative impact of noisy information and improving the robustness of the multimodal fusion. 

\subsection{Transformer}
Transformer is an attention-based building block for machine translation introduced by \cite{DBLP:conf/nips/VaswaniSPUJGKP17}. By aggregating data from the whole sequence, Transformer can learn the relationships between tokens scanned over time, replacing RNNs for a variety of tasks, such as natural language processing \citep{devlin2019bert, ding2020understanding}, computer vision \citep{DBLP:journals/corr/abs-2202-10108, DBLP:journals/corr/abs-2109-08409}, as well multimodal emotion recognition \citep{DBLP:conf/mm/HazarikaZP20, DBLP:conf/mm/LiangLJ20, DBLP:conf/acl/TsaiBLKMS19, DBLP:conf/icassp/HuangTLLN20, DBLP:conf/mm/YuanLXY21, DBLP:conf/mm/0004DFWYZS22}. \cite{DBLP:conf/acl/TsaiBLKMS19} introduced the Multimodal Transformer (MulT) to address modal data misalignment and long-distance dependencies. \cite{DBLP:conf/icassp/HuangTLLN20} utilized the Transformer to fuse audio-visual information on the model level, showing the superiority of model-level fusion over other layers of fusion strategies. \cite{DBLP:conf/mm/YuanLXY21} proposed a Transformer-based feature reconstruction network to achieve more robust multimodal emotion recognition. Despite the progress, existing Transformer-based methods mainly consider the feature from a specific modality as the query, which could introduce unnecessary noises or trivial information related to the modality. 

\begin{figure*}[ht]
	\centering
	\includegraphics[width=1.0\linewidth]{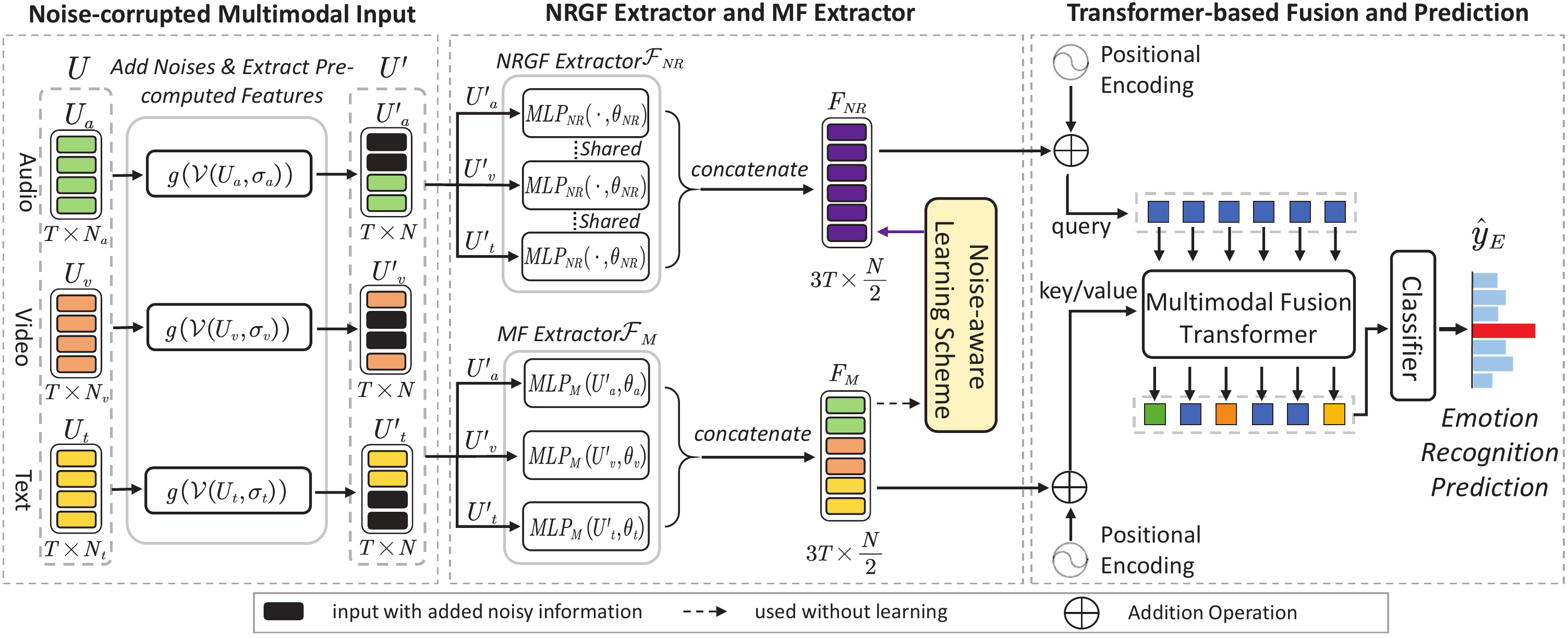}
	\caption{Processing pipeline of the proposed NORM-TR for multimodal emotion recognition. With the noise-corrupted multimodal input $U'$, we first apply a noise-resistant generic feature (NRGF) extractor to extract a generic and disturbance-insensitive representation. Then, we employ a multimodal fusion Transformer to improve NRGFs based on the more detailed multimodal features (MFs). Moreover, we introduce a novel noise-aware learning scheme to properly train the NORM-TR, thus obtaining an end-to-end noise-resistant model for emotion understanding}
	\label{fig: NORM_TR}
\end{figure*}

\section{Method}
\subsection{Overview}
The overall processing pipeline of the proposed Noise-Resistant Multimodal Transformer (NORM-TR) for the robust emotion recognition is shown in Fig.~\ref{fig: NORM_TR}. We make the NORM-TR first extract Noise-Resistant Generic Features (NRGFs) and Multimodal Features (MFs) from the input. Then, a multimodal fusion Transformer is employed to integrate the MFs according to their relations to the NRGFs, thus obtaining an end-to-end noise-resistant model for emotion understanding. To train the NORM-TR properly, we introduce a noise-aware learning scheme. In our learning scheme, we manually erase some certain periods of information in the multimodal input to implement the explicit inclusion of noisy information, which does not contain any useful semantics. By explicitly adding noisy information to either all the multimodal inputs or the input data from a specific modality, we devise two adversarial learning objectives to make the NORM-TR robust against both types of added noisy information. 

Formally, our NORM-TR employs an NRGF extractor, denoted as $\mathcal{F}_{NR}$, and an MF extractor, denoted as $\mathcal{F}_M$, to extract detailed NRGFs and MFs from its input $U'$, respectively. Then, a multimodal fusion Transformer, ${Trans}(\cdot)$, translates the MFs to the desired outputs $\hat{y}_E$ for emotion recognition according to the NRGFs. 
Therefore, our overall pipeline can be described by:
\begin{equation}
	\hat{y}_E = {Trans}\Big(\mathcal{F}_{NR}(U'), \mathcal{F}_M(U')\Big),
	\label{eq:main}
\end{equation}
where $\hat{y}_E$ is the emotion recognition output, and the $U'$ is the noise-corrupted multimodal input. It is worth mentioning that our NORM-TR, as described in Eq. \ref{eq:main}, also works for the normal multimodal input without explicitly added noises.

\subsection{Noise-corrupted Multimodal Input}
Regarding the normal multimodal emotion recognition, we use the symbol $U$ to represent the provided input multimodal information of a sequence. The $U$ can represent the input of audio, video, text, etc. In the rest of paper, we use the $U_a$, $U_v$, and $U_t$ to represent the audio, video, and text, respectively.
In the related literature \citep{DBLP:conf/acl/TsaiBLKMS19, DBLP:conf/mm/HazarikaZP20, DBLP:conf/acl/MaoYXYLG22}, \textit{pre-computed features} rather than raw data of different modalities are commonly used, thus, to be fair, the $U$ in our paper represents the pre-computed feature vectors. For example, rather than using 2D images of a video, we can have as input the pre-computed features $U_v \in \mathbb{R}^{T \times N_v}$ where $T$ represents the length of the video, and $N_v$ represents the length of the feature vector. Correspondingly, for audio and text modalities, we can also have as input the pre-computed features $U_a \in \mathbb{R}^{T \times N_a}$ and $U_t \in \mathbb{R}^{T \times N_t}$, respectively, where $N_a$ and $N_t$ are the lengths of the corresponding feature vectors, and $T$ is the unified the length of the feature vector of each modality. We would like to mention that using pre-computed features is widely accepted in the literature on multimodal emotion recognition, and the research on more appropriate features to describe multimodal raw data is beyond the scope of this study. With the pre-computed features $U=\{U_a, U_v, U_t\}$, we employ three fully-connected layers to unify the lengths of different feature vectors, respectively. We use the symbol $g$ to represent these fully-connected layers. After $g$, each modality of the obtained $U'=\{U'_a, U'_v, U'_t\}$ would have a dimension of $T\times N$, where $N$ is the unified length of feature vectors. In the meantime, we use the $\mathcal{V}(U, \sigma)$ to represent the process that explicitly adds the noisy information $\sigma$ to the input $U$. Therefore, the noise-corrupted multimodal input to our NORM-TR can be formulated as:
\begin{equation}
	U'= g(\mathcal{V}(U, \sigma)).
	\label{eq:noise}
\end{equation}

The detailed formulation of our proposed NORM-TR, such as the NRGF extractor, the MF extractor, the multimodal fusion Transformer, and the noise-aware learning scheme, will be described in the following sections, subsequently. 

\subsection{NRGF Extractor and MF Extractor}
\subsubsection{\textbf{NRGF Extractor}} With multimodal input $U'$, we introduce an NRGF extractor to obtain a generic feature that is insensitive and invariant to noisy information. We achieve this by simply employing a multi-layer perceptron:
\begin{equation}
	\label{eq:nr}
	F_{NR}=\mathcal{F}_{NR}(U') = MLP_{NR}(U',\theta_{NR}), 
\end{equation}
where $F_{NR}$ is the obtained NRGF, $MLP_{NR}$ represents multi-layer perceptrons with shared parameters for extracting NRGF. 
In practice, $MLP_{NR}$ is designed as a 2-layer fully connected network, with each layer followed by Leaky-ReLU \citep{maas2013rectifier} as the activation function. Using the $MLP_{NR}$, we reduce the feature dimension of $U'$. More specifically, if considering 3 modalities, the $U'$ obtained by Eq. \ref{eq:noise} has the dimension of $3T\times N$, and the $F_{NR}$ will have a dimension of $3T\times \frac{N}{2}$. In fact, we found that reducing the feature dimension can save parameters and achieve higher efficiency without sacrificing performance.

To make the NRGFs effective for extracting noise-insensitive features, we introduce the noise-aware learning scheme, which will be described in detail in the Section~\ref{sec: noise-aware learning scheme}.  

\subsubsection{\textbf{MF Extractor}} In addition to the NRGF extractor, we also introduce the MF extractor to extract and exploit detailed multimodal features from the input. This is because we found that the NRGF may lose some details that are beneficial for accurate emotion understanding. To implement the MF extractor, we still use a multi-layer perceptron, thus we have:
\begin{equation}
	\label{eq:mf}
	F_M=\mathcal{F}_M(U') = MLP_M(U',\theta_{*}).
\end{equation}
$MLP_M$ represents three separated multi-layer perceptrons for MF extraction, where each multi-layer perceptron architecture in $MLP_M$ is the same as $MLP_{NR}$. $*\in\{a,v,t\}$. Therefore, given 3 modalities as input, the $F_M$ will have the size of $3T \times \frac{N}{2}$.
We would like to mention that the input feature $U'$ is shared with both NRGF extractor and MF extractor. In addition, we do not make the $\mathcal{F}_M$ learn to become consistent with noise-caused changes, making it more sensitive to the variations in the multimodal input.  

\subsection{Transformer Structure and Emotion Recognition Output}
With the obtained NRGFs and MFs, we employ a multimodal fusion Transformer to achieve effective multimodal emotion recognition. The Transformer can model the relations between NRGFs and MFs, which can make MFs more appropriately complement the NRGFs based on their relevance to the NRGFs. We found that this relation modeling is important because the MFs could be more affected by noisy information, and directly fusing them with the NRGFs would introduce the negative impacts of noises (with more details in \ref{Effects of Different Components}).

\subsubsection{\textbf{Multimodal Fusion Transformer}} According to the definition of \cite{DBLP:conf/nips/VaswaniSPUJGKP17}, a Transformer takes as input the query, key, and value tensors. Then, it uses the query tensor as a reference and transforms the value tensor into desired output based on the relations between the query tensor and key tensor. When transforming, multihead attention mechanisms are performed to achieve the relation modeling and data fusion. For more details, we refer readers to \cite{DBLP:conf/acl/TsaiBLKMS19}. 

In our study, using the extracted NRGFs as query and MFs as key and value, we follow the typical formulation of the Transformer structure for more effective multimodal fusion, and implement the  ${Trans(\cdot)}$ as:
\begin{equation}
	Trans(\cdot) = Trans(q=F_{NR}, k/v=F_M),
\end{equation}
where $q$, $k$, and $v$ represent the query, key, and value tensors in a Transformer, respectively. 
More specifically, the query, \textit{i.e.,} NRGFs extracted by $\mathcal{F}_{NR}$, have a shape of $3T \times \frac{N}{2}$. The key and value tensors share the same multimodal feature obtained by $\mathcal{F}_{M}$, which have a shape of $3T \times \frac{N}{2}$. 
The $Trans(\cdot)$ contains eight attention heads and attention blocks of different depths on different datasets, \textit{e.g.,} 2 levels of depth for MOSI
\citep{DBLP:journals/expert/ZadehZPM16} and RML \citep{DBLP:journals/tmm/WangG08} datasets, and 4 levels of depth for MOSEI \citep{DBLP:conf/acl/MorencyCPLZ18} and IEMOCAP \citep{DBLP:journals/lre/BussoBLKMKCLN08} datasets. This is because using more depths for smaller datasets like MOSI and RML can easily result in over-fitting. In addition, following~\cite{DBLP:conf/iclr/DosovitskiyB0WZ21}, we perform a learnable positional encoding of timestamp of the sequence and add it to the input of the Transformer. It is worth mentioning that we have tried more complicated structures for fusion, but it does not improve performance quite much (with more details in ~\ref{sec: Effects of the Hyper-parameter Settings in Transformer}). We found that deep Transformer structures can also overfit the dataset easily. 

\subsubsection{\textbf{Emotion Recognition Output}}
After the multimodal fusion Transformer, we obtain the final emotion recognition output $\hat{y}_E$ by applying an emotion classification layer on the outputs of Transformer. Specifically, the Transformer outputs a tensor of shape $3T \times \frac{N}{2}$ after relation modeling and tensor fusion. Then, we apply an average pooling operation on the obtained tensor to reduce its dimension from $3T \times \frac{N}{2}$ to $1 \times \frac{N}{2}$ for relieving the computational burden. The pooled tensor is later fed into a fully connected layer for emotion classification. Lastly, we have the $\hat{y}_E$ of a shape $1 \times C$, where $C$ represents the number of categories. Each element of the $\hat{y}_E$ represents a specific emotion like happy and angry. In general, with the help of the NRGF extractor and the Transformer, we obtain an accurate estimation $\hat{y}_E$ that is much less affected by noisy information. 

\subsection {Noise-aware Learning Scheme} 
\label{sec: noise-aware learning scheme}
By devising the NORM-TR, it is essential to apply appropriate learning objectives to help our model learn to resist noisy information. Therefore, we introduce a novel noise-aware learning scheme for training the NORM-TR and making it robust to noises. Our novel learning scheme explicitly adds two types of noisy information to corrupt raw multimodal data and then introduces two corresponding adversarial losses to encourage the NORM-TR to provide an NRGF invariant to the added noisy information. 

\subsubsection{\textbf{Explicit Noisy Information}} 
\begin{figure}[htbp]
	\centering
	\includegraphics[width=1.0\linewidth]{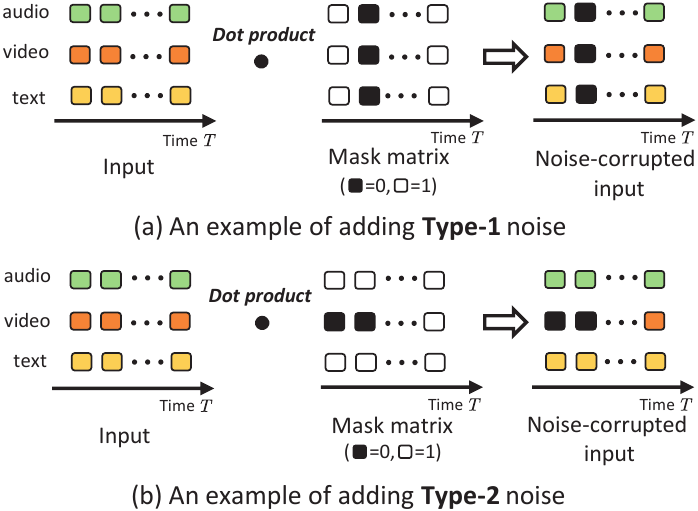}
	\caption{Examples of adding noises of two types. (a) Adding \textbf{Type-1} noise, (b) adding \textbf{Type-2} noise}
	\label{fig: noise_conception}
\end{figure}

We explicitly add noisy information as described in Eq. \ref{eq:noise} because it is extremely difficult to properly define what patterns should belong to noisy information and what should not. Without explicit noisy information, we would not know when a model should be insensitive to the changes of patterns in the input sequence. Therefore, we attempt to explicitly add noisy information to the multimodal input. Regarding this, we randomly erase some certain periods of input multimodal representation to remove any potential semantics in the input data. In this study, we mainly consider two types of noisy information: (\textbf{Type-1}) all the input multimodal data contains random noisy information (see Fig.~\ref{fig: noise_conception}(a)); and (\textbf{Type-2}) only the input data of a specific modality contains random noisy information (see Fig.~\ref{fig: noise_conception}(b)). Training on the first case can help the NORM-TR summarize generic and globally consistent semantics, and the second case can help the NORM-TR focus on improving its robustness against noisy information to each specific modality. Our experimental results can validate the importance of adding both types of noisy information. In practice, we masking-out input information to implement the Eq.\ref{eq:noise}. For example, when adding the Type-1 noise, we generate a mask for the pre-computed multimodal feature vector of each modality. In each mask, we randomly sample a time window whose length ranges from 0 to $\frac{T}{2}$ time steps, and then we set the values within this time window to 0 with the remaining values of this mask being 1. Thus, each modality has its individual mask for processing. Then, we multiply these masks with the pre-computed multimodal feature vectors $U$, erasing the semantics contained in the period corresponding to the sampled time window of this mask. In addition, for adding the Type-2 noise, the semantic erasing operations are similar to the procedure of adding the Type-1 noise, but the mask generation is different. In this case, we only generate one mask, in which a time window containing 0s is sampled randomly. We multiply this mask to the pre-comupted feature of a random modality, thus adding the noise to this modality only. 
\begin{figure}[htbp]
	\centering
	\includegraphics[width=1.0\linewidth]{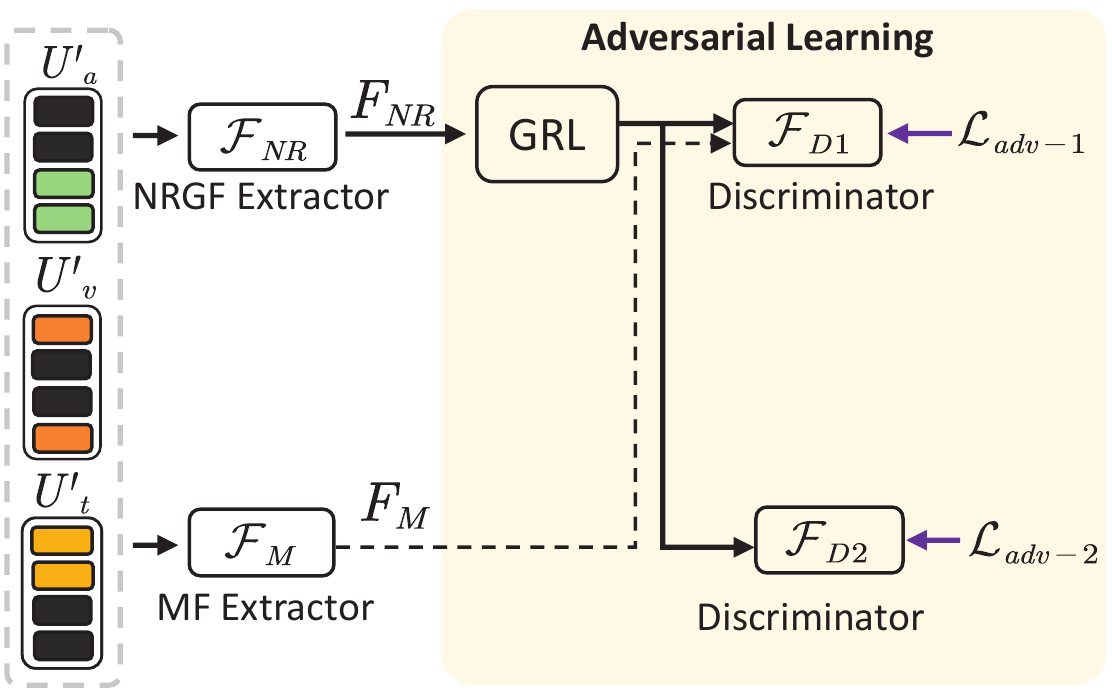}
	\caption{The pipeline of the noise-aware learning scheme}
	\label{fig: Noise-aware Learning Scheme}
\end{figure}

As shown in Fig.\ref{fig: Noise-aware Learning Scheme}, after adding noisy information, we use two adversarial loss functions to define learning objectives regarding both types of added noisy information, respectively. We apply adversarial loss functions because they are powerful at making a model produce the high-dimensional output with some desired patterns, \textit{e.g.,} producing a fake 2D image that has very similar image patterns to the desired real 2D images. To implement the adversarial-based learning, a discriminator is employed to learn to distinguish whether the generated output fails to obtain the desired patterns. The unsuccessfully generated output will be easily identified by the discriminator, while the successfully generated output will confuse the discriminator. The detailed generative adversarial learning scheme can be found in \cite{DBLP:conf/nips/GoodfellowPMXWOCB14}.

\subsubsection{\textbf{Adversarial Learning-based Learning Scheme}}
Here, we attempt to use adversarial losses to make our NORM-TR learn to produce the NRGF that is invariant to noises. By resisting the impacts of the added noisy information via adversarial learning, the NRGF extractor would be more effective at extracting semantically meaningful features from the input data, which can further improve the robustness of emotion recognition. Formally, we use the symbols $\mathcal{F}_{D1}$, $\mathcal{F}_{D2}$ to represent the discriminators of Type-1 and Type-2 added noises, respectively. Fig. \ref{fig: Noise-aware Learning Scheme} shows the detailed adversarial learning structure of our noise-aware learning scheme. 

More specifically, since the Type-1 noises are randomly added to all the multimodal input, the NRGF extractor is supposed to find generic and universal patterns from the noise-corrupted multimodal input, so that the extracted NRGF can be invariant to the Type-1 noises.
To achieve this, we employ a discriminator $\mathcal{F}_{D1}$ and make it learn to classify whether its input feature contains modality-dependant patterns: if the input to  $\mathcal{F}_{D1}$ is a modality-specific feature from the MFs $F_M$, we make this discriminator learn to predict which modality this feature represents; if the input to $\mathcal{F}_{D1}$ is the NRGF, we make the NRGF extractor confuse this discriminator. Therefore, when the NRGF extractor fails to provide generic and consistent features, it would not confuse the discriminator effectively. By confusing the discriminator, the extracted NRGF would be representative for the entire multimodal input rather than a specific modality that can be more likely to be affected by the added Type-1 noise. To sum up, suppose the model parameter of NRGF extractor is $\theta_{NR}$, and the model parameters of the discriminator $\mathcal{F}_{D1}$ are $\theta_{D1}$. Then, the first objective of our noise-aware learning scheme can be described by the optimization \textit{w.r.t.} the adversarial losses $\mathcal{L}_{adv-1}$ for Type-1 added noisy information:

\begin{footnotesize}
	\begin{equation}
		\min_{\substack{\theta_{D1}}}
		\max_{\theta_{NR}} \mathcal{L}_{adv-1} = 
		-\frac{1}{N_{b}} \sum_{i=0}^{N_{b}}
		y^{i}_{M} \cdot log ~ \mathcal{F}_{D1}(F_{NR}/F_{M}; \theta_{D1}),
		\label{eq: L_adv-1}
	\end{equation}
\end{footnotesize}
where $N_b$ is the number of samples in the training set,  $y^{i}_M$ represents the label indicating which modality the $F_{NR}$ or $F_{M}$ comes from, and $F_{NR}$ is the NRGF extracted according to the parameter $\theta_{NR}$. 
To make the NRGF compatible to $\mathcal{F}_{D1}$ that identifies the modality-dependent features, we add 3 extra MLPs to explain the $F_{NR}$ of $3T \times \frac{N}{2}$ into the features related to 3 input modalities, respectively, with a shape of $3\times (T \cdot \frac{N}{2})$. 

Regarding the Type-2 noises, we also employ a discriminator $\mathcal{F}_{D2}$ to help train the NRGF extractor. Different from the Type-1 noises, we add noises to only one random modality. Learning against the Type-2 noises can help the NORM-TR works effectively on the input without adding explicit noises while still maintaining its capability of being robust against noises existing in a random modality. Regarding the Type-2 noises, we simply make the discriminator $\mathcal{F}_{D2}$ identify whether a modality contains noises: if the input to  $\mathcal{F}_{D2}$ is related to the input without added noises, we make this discriminator predict a negative label; otherwise, we make this discriminator predict a positive label. Therefore, if the NRGF fails to be invariant to the noises only added to a specific modality, the NRGF extractor would also fail to confuse the discriminator and the $\mathcal{F}_{D2}$ can easily identify the noises by predicting a positive label. To this end, suppose the model parameters of the $\mathcal{F}_{D2}$ are $\theta_{D2}$. We have the second objective of our noise-aware learning scheme as the optimization \textit{w.r.t.} the adversarial losses $\mathcal{L}_{adv-2}$ for Type-2 added noisy information:
\begin{small} 
	\begin{equation}
		\min_{\substack{\theta_{D2}}}
		\max_{\theta_{NR}} \mathcal{L}_{adv-2}= 
		-\frac{1}{N_{b}} \sum_{i=0}^{N_{b}}
		y^{i}_{N} \cdot log ~ \mathcal{F}_{D2}(F_{NR}; \theta_{D2}),
		\label{eq: L_adv-2}
	\end{equation}
\end{small}
where $y^i_N$ is the label indicating which modality is corrupted by added noises.

In practice, we use the fully-connected layer to implement the two discriminators, each of which consists of a fully-connected layer. In addition, we apply the gradient reversal layer (GRL)~\citep{DBLP:conf/icml/GaninL15} to implement the adversarial learning \textit{w.r.t.} NRGF extractor that is supposed to confuse the two discriminators. 

\subsection{Overall Learning Objectives}
To sum up, our method involves three learning objectives, including two adversarial loss functions $\mathcal{L}_{adv-1}$ and $\mathcal{L}_{adv-2}$, and one final emotion learning loss $\mathcal{L}_{er}$. Considering that the emotion labels on different datasets are different, for example, the labels on the RML \citep{DBLP:journals/tmm/WangG08} and IEMOCAP \citep{DBLP:journals/lre/BussoBLKMKCLN08} datasets are discrete, while the labels on the MOSI \citep{DBLP:journals/expert/ZadehZPM16} and MOSEI \citep{DBLP:conf/acl/MorencyCPLZ18} datasets are continuous. Therefore, we introduce the cross-entropy loss as the emotion learning loss $\mathcal{L}_{er}$ for classification on the RML and IEMOCAP datasets, and the mean squared error (MSE) as $\mathcal{L}_{er}$ for regression on the MOSI and MOSEI datasets.
The emotion learning loss $\mathcal{L}_{er}$ can be written as: 

\begin{small}
	\begin{equation}
		\mathcal{L}_{er} = \left\{ \begin{array}{c}
			-\frac{1}{N_{b}} \sum_{i=0}^{N_{b}} y^{i} \cdot \log \hat{y}^{i}_E  \quad \text{for classification}  \\
			\\
			\frac{1}{N_b} \sum_{i=0}^{N_b}\left\|{y}^i-\hat{{y}}^i_E\right\|_2^2   \quad \text{for regression} 
		\end{array}\right. 
	\end{equation}
\end{small}
where $y^{i}$ is the emotion label of the $i$-th sample. $\hat{y}^{i}_E$ is the prediction of NORM-TR. 
The overall objective function $\mathcal{L}$ is the sum of $\mathcal{L}_{adv-1}$, $\mathcal{L}_{adv-2}$, and $\mathcal{L}_{er}$. Mathematically, the $\mathcal{L}$ can be written as:  
\begin{small} 
	\begin{equation}
		\mathcal{L} = \mathcal{L}_{adv-1} + \mathcal{L}_{adv-2} + \mathcal{L}_{er}.
	\end{equation}
\end{small}

\section{Experiment} \label{sec:Exp}
\subsection{Datasets}
We conducted extensive experiments on three trimodal datasets (MOSI~\citep{DBLP:journals/expert/ZadehZPM16}, MOSEI~\citep{DBLP:conf/acl/MorencyCPLZ18}, and IEMOCAP~\citep{DBLP:journals/lre/BussoBLKMKCLN08}), as well as a bimodal emotion dataset (RML~\citep{DBLP:journals/tmm/WangG08}). 
These datasets cover different languages and various scenarios like natural and laboratory scenes, dialogue, and solo presentations.

\textbf{MOSI}
The MOSI dataset consists of 2,199 multimodal sequence samples with video, audio, and text modalities. The training, validation, and testing sets of MOSI contain 1,284 samples, 229 samples, and 686 samples, respectively. Each multimodal sample has a uniform label that ranges from -3 to 3. The -3 and +3 represent strongly negative and strongly positive emotions, respectively.

\textbf{MOSEI}
The MOSEI dataset comprises 22,851 video clips collected from YouTube with spontaneous expressions, head poses, occlusions, illuminations, and so on. This dataset is divided into 16,326 training samples, 1,871 validation samples, and 4,659 test samples in speaker-independent settings. Each sample is manually annotated with a sentiment score from -3 (strongly negative) to 3 (strongly positive).

\textbf{IEMOCAP}
The IEMOCAP includes video, audio, and text transcriptions and 12 hours of video recordings of situational dialogues. The data is divided into five sessions with a total of 10,039 samples and 9 emotion categories. Following the comparison methods \citep{DBLP:conf/acl/TsaiBLKMS19, DBLP:conf/cvpr/LvCHDL21}, we used the four emotion categories, \textit{i.e.}, happiness, anger, sadness, and neutral. The data is partitioned into 2,717 training samples, 798 validation samples, and 938 testing samples. 

\textbf{RML}
The RML is an audio-visual emotion dataset, containing 8 topics with 720 video samples, with six different languages (\textit{i.e.}, English, Mandarin, Urdu, Punjabi, Persian, and Italian). Each data was labeled as one of six emotions, \textit{i.e.,} anger, disgust, fear, happiness, sadness, and surprise. The training and testing sets were in a ratio of about 8:2 for cross-validation with speaker-independent settings, which ensures that the speakers in the training set were not in the corresponding test set.

\subsection{Implementation Details}
\label{sec: Implementation Details}
We used PyTorch to implement our method. The experiments were conducted on a PC with Intel(R) Xeon(R) Gold 6240C CPU at 2.60GHz and 128GB memory and NVIDIA GeForce RTX 3090. The key training parameters include initial learning rate (0.0001), cosine annealing schedule to adjust the learning rate, mini-batch size (16), and warm up. 

For the sequence length setting, we unified the length of the sequences to 8 on the RML and IEMOCAP datasets (\textit{i.e.}, $T=8$), and 50 on the MOSI and MOSEI datasets (\textit{i.e.}, $T=50$). The generation of video modality is different from audio and text modalities. More specifically, for the video modality, we divided the input video into $T$ segments and randomly sampled one frame from each of the segment to form a video sequence of length $T$. For the audio and text modalities, we directly  truncated the first $T$ frames of the data as the input sequence.

\subsection{Pre-computed Feature Extraction}
\label{sec: Embedding Feature Extraction}
\textbf{Video features $U_v$:}
For the RML and IEMOCAP datasets, following the existing method \citep{DBLP:conf/aaai/ZhaoMGYXXHCK20}, we employed a ResNet-18 \citep{DBLP:journals/corr/HeZRS15} to extract the last global averaging pooling output of the ResNet-18 as the pre-computed video features. For the MOSI and MOSEI datasets, referring to the existing methods \citep{DBLP:conf/acl/TsaiBLKMS19, DBLP:conf/mm/HazarikaZP20, DBLP:conf/acl/MaoYXYLG22}, we used the features provided in the dataset, which had been extracted by the OpenFace~\citep{DBLP:conf/wacv/Baltrusaitis0M16}. 

\textbf{Audio features $U_a$:}
For RML and IEMOCAP, we first used Librosa to compute the log mel-spectrogram and its first and second-order differentials of each sample, and then employed a ResNet-18 \citep{DBLP:journals/corr/HeZRS15} to extract  features. Finally, we stacked all feature vectors and obtained the pre-computed audio features. For MOSI and MOSEI, we used the features provided by the dataset, which were extracted by the Librosa \footnote{https://librosa.org}. 

\begin{table*}[!ht]
	\begin{minipage}[t]{\textwidth}
		\small
		\caption{Comparison results on MOSI dataset}
		\label{tab:Results_Compare_MOSI}
		\begin{tabular*}{\linewidth}{l@{\extracolsep{\fill}}ccccc}
			\toprule
			Methods & Acc-7 ($\uparrow$) & Acc-2 ($\uparrow$) & F1 ($\uparrow$) & MAE ($\downarrow$) & Corr ($\uparrow$) \\
			\midrule
			TFN \citep{DBLP:conf/emnlp/ZadehCPCM17} & 34.9 & - / 80.8 & - / 80.7 & 0.901 & 0.698 \\
			LMF \citep{DBLP:conf/acl/MorencyLZLSL18} & 33.2 & - / 82.5 & - / 82.4 & 0.917 & 0.695 \\
			MFM \citep{DBLP:conf/iclr/TsaiLZMS19} & 35.4 & - / 81.7 & - / 81.6 & 0.877 & 0.706 \\
			ICCN \citep{DBLP:conf/aaai/SunSSL20} & 39.0 & - / 83.0 & - / 83.0 & 0.877 & 0.706 \\
			MuLT \citep{DBLP:conf/acl/TsaiBLKMS19} & 40.0 & - / 83.0 & - / 82.8 & 0.871 & 0.698 \\
			MISA \citep{DBLP:conf/mm/HazarikaZP20} & 42.3 & 81.8 / 83.4 & 81.7 / 83.6 & 0.783 & 0.761 \\
			PMR \citep{DBLP:conf/cvpr/LvCHDL21} & 40.6 & - / 83.6 & - / 83.4 & - & - \\
			Self-MM \citep{DBLP:conf/aaai/YuXYW21} & 45.8 & 84.0 / 86.0 & 84.4 / 86.0 & 0.713 & 0.798 \\
			FDMER \citep{DBLP:conf/mm/YangHKDZ22} & 44.1 & - / 84.6 & - / 84.7 & 0.724 & 0.788 \\
			\textbf{Our NORM-TR} & \textbf{48.5} & \textbf{84.3} / \textbf{86.1} & \textbf{84.4} / \textbf{86.2} & \textbf{0.698} & \textbf{0.808} \\
			\bottomrule
		\end{tabular*}
		\newline 
		Note: for each evaluation metric, $\uparrow$ indicates the bigger the better while $\downarrow$ indicates the smaller the better. The best result is highlighted in bold
	\end{minipage}
\end{table*}

\textbf{Text features $U_t$:}
For IEMOCAP, we used a pre-trained BERT \citep{devlin2019bert} as the feature extractor to encode the transcribed word sequences into the pre-computed text features. For MOSI and MOSEI datasets, we also used the text features provided by the dataset, which were extracted by BERT.  

\begin{table*}[!ht]
	\begin{minipage}[t]{\textwidth}
		\small
		\caption{Comparison results on MOSEI dataset}
		\label{tab:Results_Compare_MOSEI}
		\begin{tabular*}{\linewidth}{l@{\extracolsep{\fill}}ccccc}
			\toprule
			Method & Acc-7 ($\uparrow$) & Acc-2 ($\uparrow$) & F1 ($\uparrow$) & MAE ($\downarrow$) & Corr ($\uparrow$) \\
			\midrule
			TFN~\citep{DBLP:conf/emnlp/ZadehCPCM17} & 50.2 & - / 82.5 & - / 82.1 & 0.593 & 0.700 \\
			LMF~\citep{DBLP:conf/acl/MorencyLZLSL18} & 48.0 & - / 82.0 & - / 82.1 & 0.623 & 0.677 \\
			MFM ~\citep{DBLP:conf/iclr/TsaiLZMS19} & 51.3 & - / 84.4 & - / 84.3 & 0.568 & 0.717 \\
			ICCN~\citep{DBLP:conf/aaai/SunSSL20} & 51.6 & - / 84.2 & - / 84.2 & 0.565 & 0.713 \\
			MuLT~\citep{DBLP:conf/acl/TsaiBLKMS19} & 51.8 & - / 82.5 & - / 82.3 & 0.580 & 0.703 \\
			MISA~\citep{DBLP:conf/mm/HazarikaZP20} & 52.2 & 83.6 / 85.5 & 83.8 / 85.3 & 0.555 & 0.756 \\
			PMR~\citep{DBLP:conf/cvpr/LvCHDL21} & 52.5 & - / 83.3 & - / 82.8 & - & - \\
			Self-MM~\citep{DBLP:conf/aaai/YuXYW21} & 53.5 & 82.8 / 85.2 & 82.5 / 85.3 & 0.530 & 0.765 \\
			FDMER ~\citep{DBLP:conf/mm/YangHKDZ22} & 54.1 & - / 86.1 & - / 85.8 & 0.536 & 0.773 \\
			\textbf{Our NORM-TR} & \textbf{54.6} & \textbf{84.3} / \textbf{86.6} & \textbf{84.5} / \textbf{86.6} & \textbf{0.529} & \textbf{0.778} \\
			\bottomrule
		\end{tabular*} 
		\newline 
		Note: for each evaluation metric, $\uparrow$ indicates the bigger the better while $\downarrow$ indicates the smaller the better. The best result is highlighted in bold
	\end{minipage}
\end{table*}

\subsection{Evaluation Metrics}
On the RML dataset, we chose two widely-used evaluation metrics, \textit{i.e.,} six classification accuracy (Acc-6) and F-Score (F1) to evaluate the performance. On the IEMOCAP dataset, we followed previous works~\citep{DBLP:conf/cvpr/LvCHDL21} to report the binary classification accuracy (Acc-2) and F1 for each emotion category. 
On MOSI and MOSEI, referring to prior works \citep{DBLP:conf/acl/YuXMZMWZY20}, we used six widely-used evaluation metrics, \textit{i.e.}, Acc-2, F1, three classification accuracy (Acc-3), seven classification accuracy (Acc-7), mean absolute error (MAE), and the correlation of the model’s prediction with human (Corr). More specifically, following prior works~\citep{DBLP:conf/mm/HazarikaZP20, DBLP:conf/aaai/YuXYW21}, we calculated Acc-2 and F1 in two ways: negative/non-negative and negative/positive on MOSI and MOSEI datasets, respectively.

In addition, to validate the robustness of our method to different intensities of noise, referring to previous work \citep{DBLP:conf/mm/YuanLXY21}, we introduced Area Under Indicators Line Chart (AUILC) to evaluate the noise robustness of our method on the test set.
The AUILC can be written as:
\begin{equation}
	\text {AUILC}=\sum_{t} \frac{\left(x_t+x_{t+1}\right)}{2} \cdot\left(r_{t+1}-r_t\right)
\end{equation}
where $x_t$ and $x_{t+1}$ represent the $t$-th and $t+1$-th evaluation results under masking percentages of $r_t$ and $r_{t+1}$, respectively. 

\subsection{Overall Performance}

\begin{table*}
	\caption{Comparison results on IEMOCAP dataset}
	\label{tab:Results_Compare_IEMOCAP}
	\small
	\renewcommand\arraystretch{1.1}
	\begin{tabular*}{\linewidth}{l@{\extracolsep{\fill}}cccccccccc}
		\toprule
		\multirow{2}{*}{Methods}  & \multicolumn{2}{c}{Happiness}&  \multicolumn{2}{c}{Sadness}& \multicolumn{2}{c}{Anger}& \multicolumn{2}{c}{Neutral}& \multicolumn{2}{c}{Average} \\
		& Acc-2 & F1 & Acc-2 &F1 & Acc-2 & F1 & Acc-2 & F1  & Acc-2  & F1        \\ 
		\midrule
		EF-LSTM
		& 76.2  & 75.7&70.2&70.5&72.7&67.1&58.1&57.4&69.3&67.7   \\ 
		LF-LSTM
		& 72.5 & 71.8&72.9&70.4&68.6&67.9&59.6&56.2&68.4&66.6  \\ 
		RAVEN~\citep{DBLP:conf/aaai/WangSLLZM19}
		& 77.0 & 76.8&67.6&65.6&65.0&64.1&62.0&59.5&67.9&66.5  \\ 
		MCTN~\citep{DBLP:conf/aaai/PhamLMMP19}
		& 80.5 & 77.5&72.0&71.7&64.9&65.6&49.4&49.3&66.7&66.0   \\
		MulT~\citep{DBLP:conf/acl/TsaiBLKMS19}
		& 84.8 & 81.9&77.7&74.1&73.9&70.2&62.5&59.7&74.7&71.5 \\ 
		PMR~\citep{DBLP:conf/cvpr/LvCHDL21}
		& 86.4 & 83.3 &78.5&75.3&75.0&71.3&63.7&60.9&75.9&72.7\\
		ScaleVLAD\citep{DBLP:journals/corr/abs-2112-01368}
		& 86.7 & 85.9 &84.8&84.6&86.8&86.9&72.1&72.1&82.6&82.4\\ 
		\textbf{Our NORM-TR} 
		& \textbf{87.7} & \textbf{88.5 }&\textbf{86.2}&\textbf{86.4}&\textbf{88.6}&\textbf{88.6}&\textbf{74.8}&\textbf{74.3}&\textbf{84.3}&\textbf{84.5}\\ 
		\bottomrule
	\end{tabular*}
	Note: the best result is highlighted in bold
\end{table*}

\subsubsection{Experiments on the MOSI Dataset}
Table~\ref{tab:Results_Compare_MOSI} lists the comparison results of our proposed method and state-of-the-art methods on the MOSI dataset.  As shown in the table, the proposed NORM-TR achieved an improvement of 2.7\% on the Acc-7 compared to the second best result obtained by Self-MM~\citep{DBLP:conf/aaai/YuXYW21}. Compared to the other Transformer-based method FDMER~\citep{DBLP:conf/mm/YuanLXY21}, our method gained a relative improvement of 9.98\% on the Acc-7. Moreover, we also achieved state-of-the-art performance on all other metrics, especially on the more difficult seven classification task. We attribute such a large improvement to the fact that the extracted noise-resistant features can help our NORM-TR suppress useless noisy information during the fusion process, thus improving multimodal emotion recognition.

\subsubsection{Experiments on the MOSEI Dataset}
Table~\ref{tab:Results_Compare_MOSEI} reports the comparison results of our method and state-of-the-art methods on the MOSEI dataset. Our NORM-TR achieved significant improvement at the Acc-2, F1, Acc-3, Acc-5, MAE, and Corr. Compared to these Transformer-based methods, namely FDMER~\citep{DBLP:conf/mm/YangHKDZ22}, and MulT~\citep{DBLP:conf/acl/TsaiBLKMS19}, we achieved relative improvements in all metrics, \textit{e.g.}, with 0.92\% on Acc-7 and 0.93\% on F1, respectively. Achieving such superior performance on large-scale datasets with more complex scenarios demonstrates the ability of our NORM-TR to extract effective emotion information from various scenarios.

\subsubsection{Experiments on the IEMOCAP Dataset}
Table~\ref{tab:Results_Compare_IEMOCAP} shows the comparison results of our method and state-of-the-art methods, including MulT~\citep{DBLP:conf/acl/TsaiBLKMS19}, PMR~\citep{DBLP:conf/cvpr/LvCHDL21}, and ScaleVLAD~\citep{DBLP:journals/corr/abs-2112-01368}, on the IEMOCAP dataset. It is observed that our proposed NORM-TR achieved the best performance, which demonstrates the superiority of our NORM-TR. Compared with the state-of-the-art method ScaleVLAD, our NORM-TR achieved a relative 2.06\% and 2.55\% improvements on the averaged Acc and F1, respectively. In addition, we also achieved state-of-the-art performance for all four categories on the binary accuracy corresponding F1.

\subsubsection{Experiments on the RML Dataset}
Table~\ref{tab:Results_Compare_RML} reports the comparison results of our method and state-of-the-art methods on the RML dataset. Compared to the second best result obtained by MulT~\citep{DBLP:conf/acl/TsaiBLKMS19}, our NORM-TR achieved a relative 3.23\% boost on the averaged accuracy. Compared to the Census-Transform proposed by \cite{DBLP:conf/smc/CornejoP19}, our NORM-TR achieved a greater relative improvement of 7.75\%. It shows that our method effectively addresses the effect of noise information to improve the performance of Transformer.
\begin{table}
	\caption{Comparison results on RML}
	\label{tab:Results_Compare_RML}
	\begin{tabular*}{\linewidth}{l@{\extracolsep{\fill}}cc}
		\toprule
		Methods & Acc-6 & F1  \\
		\midrule
		Madany \textit{et al.} \citep{DBLP:conf/iscas/El-MadanyHG16} & 75.00&- \\
		Zhang \textit{et al.}\citep{DBLP:journals/tcsv/ZhangZHGT18} & 80.36 &-\\
		Ma \textit{et al.}\citep{DBLP:journals/inffus/MaHCCLK19} & 80.46 &-\\
		Cornejo \textit{et al.} \citep{DBLP:conf/smc/CornejoP19} & 82.50 &-\\
		TFN$^*$~\citep{DBLP:conf/emnlp/ZadehCPCM17} & 83.19&83.22 \\
		MulT$^*$~\citep{DBLP:conf/acl/TsaiBLKMS19} & 86.11&85.87\\
		MISA$^*$~\citep{DBLP:conf/mm/HazarikaZP20} & 81.11&80.78\\
		\textbf{Our NORM-TR} & \textbf{88.89}&\textbf{88.81}\\
		\bottomrule
	\end{tabular*}
	Note: the best result is highlighted in bold and $^*$ indicates that the result is reproduced by authors
\end{table}

\begin{table*}[!ht]
		\caption{Model robustness comparison on MOSI and MOSEI datasets}
		\label{tab:Results_Compare_Robustness}
		\small
		\resizebox{1.0\linewidth}{!}
		{
			\begin{tabular}{lcccccccc}
				\toprule
				\multirow{2}{*}{Method} & \multicolumn{4}{c}{MOSI} & \multicolumn{4}{c}{MOSEI} \\
				& Acc-7 ($\uparrow$) & Acc-2 ($\uparrow$) & MAE ($\downarrow$) & Corr ($\uparrow$) & Acc-7 ($\uparrow$) & Acc-2 ($\uparrow$) & MAE ($\downarrow$) & Corr ($\uparrow$) \\
				\midrule
				TFN~\citep{DBLP:conf/emnlp/ZadehCPCM17} & - & - / 60.4 & 1.327 & 0.300
				& - & - / - & - & - \\
				MulT~\citep{DBLP:conf/acl/TsaiBLKMS19} & - & - / 61.8 & 1.288 & 0.334
				& - & - / - & - & - \\
				MISA~\citep{DBLP:conf/mm/HazarikaZP20} & - & - / 63.2 & 1.209 & 0.403
				& - & - / - & - & - \\
				TFR-Net~\citep{DBLP:conf/mm/YuanLXY21} & - &  - / \textbf{69.0} & 1.155 & 0.467 & - & - / - & - & - \\
				TFN*~\citep{DBLP:conf/emnlp/ZadehCPCM17} & 22.0 & 62.6 / 62.6 & 1.281 & 0.342
				& 45.1 & 70.8 / 72.6 & 0.726 & 0.409 \\
				MulT*~\citep{DBLP:conf/acl/TsaiBLKMS19} & 22.7& 63.5 / 64.0 & 1.265& 0.339
				& 46.4 & 74.5 / 75.5 & 0.692& 0.504 \\
				MISA*~\citep{DBLP:conf/mm/HazarikaZP20} & 27.0 & 65.4 / 65.4 & 1.181 & 0.412
				& 44.5 & 73.3 / 74.5 & 0.720 & 0.421 \\
				TFR-Net*~\citep{DBLP:conf/mm/YuanLXY21} &26.8 &  67.8 / 68.2 & 1.175 & 0.445 & 46.5 & 73.3 / 74.7 & 0.686 & 0.515 \\
				\textbf{Our NORM-TR} & \textbf{31.3} & \textbf{68.6} / {68.6} & \textbf{1.093} & \textbf{0.506} & \textbf{47.3} &      \textbf{74.8} / \textbf{76.5} & \textbf{0.673} & \textbf{0.549} \\
				\bottomrule
			\end{tabular}
		}
		\newline 
		Note: for each evaluation metric, $\uparrow$ indicates the bigger the better while $\downarrow$ indicates the smaller the better. The best result is highlighted in bold, $^*$ indicates that the result is reproduced by authors
\end{table*}

\begin{figure*}[!ht]
	\centering
	\includegraphics[width=1.0\textwidth]{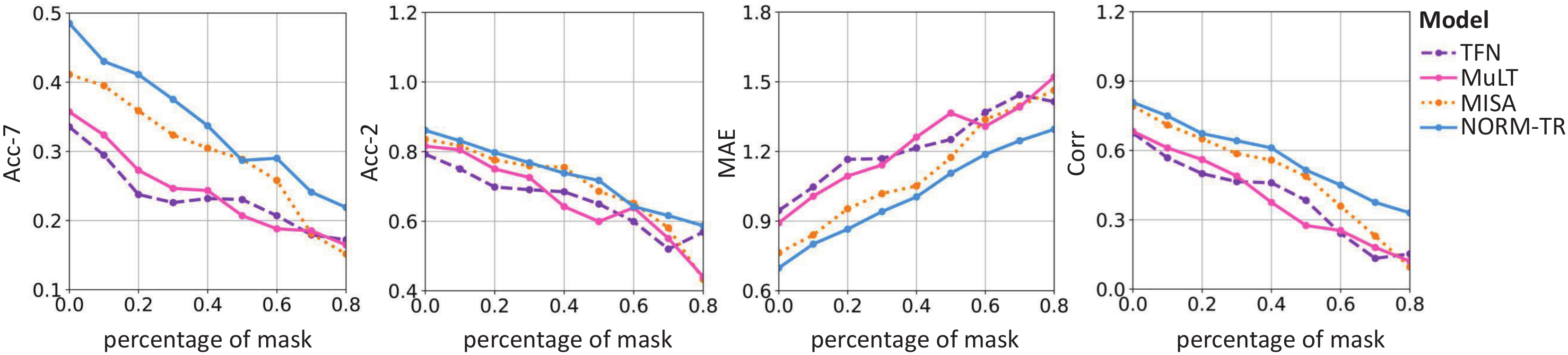}
	\caption{Visualization of the metrics curves for test data under various mask percentages on the MOSI dataset}
	\label{fig: metric_curves}
\end{figure*}

\subsection{Robustness Evaluation for Noisy Data}
To further verify the robustness of our method to noisy data, we evaluated our method with test data under different masking percentages $r_t$, where $r_t \in\{0, 0.1, ... ,1.0 \}$. Table~\ref{tab:Results_Compare_Robustness} reports the AUILC results of our method and state-of-the-art methods on the MOSI and MOSEI datasets, respectively. It is obvious that our NORM-TR achieved better performance on almost all metrics on the MOSI dataset, \textit{i,e.,} 31.3\% on Acc-7, 68.6\% on Acc-2, 1.093 on MAE, and 0.506 on Corr, respectively. On the MOSEI dataset, our method also obtained significant improvements on all metrics. For example, compared to TFR-Net~\citep{DBLP:conf/mm/YuanLXY21}, our method obtained a relative improvement of 1.72\% on the Acc-7, 2.05\% / 2.41\% on the Acc-2, 1.90\% on the MAE, and 6.60\% on the Corr, respectively. It demonstrates the great robustness of our method in the face of noise disturbances.

In addition, Fig.~\ref{fig: metric_curves} shows the metric curves with the test data under various mask percentages on MOSI dataset. As shown in the figure, NORM-TR outperforms the other methods on almost all evaluation metrics at various mask percentages $r_t \in \{0, 0.1,  \cdots, 0.8 \}$, indicating that our NORM-TR achieves greater robustness to noisy.

\begin{table*}[htpb]
	\small
	\caption{Ablation study of the proposed NORM-TR. Impact of subtraction of the different components into the NORM-TR on the MOSI and MOSEI dataset, respectively}
	\label{tab: Effects of different Components}
	\begin{tabular*}{\linewidth}{c@{\extracolsep{\fill}}cccc}
		\toprule
		\multirow{2}{*}{Method} & \multicolumn{2}{c}{MOSI} & \multicolumn{2}{c}{MOSEI} \\
		& Acc-7 ($\uparrow$) & MAE ($\downarrow$) & Acc-7 ($\uparrow$) & MAE ($\downarrow$) \\ 
		\midrule
		\textbf{NORM-TR} &\textbf{48.5}&\textbf{0.698} & \textbf{54.6} & \textbf{0.528} \\ 
		w/o NRGF Extractor & 46.8 & 0.726 & 52.3 & 0.547 \\
		w/o MF Extractor & 48.1 & 0.706 & 53.6 & 0.534 \\
		w/o Multimodal fusion Transformer & 45.6 & 0.734 & 53.8 & 0.536 \\
		w/o Noise-aware Learning Scheme & 46.9 & 0.721 & 53.2 & 0.544 \\
		\bottomrule
	\end{tabular*}
	\newline
	Note: for each evaluation metric, $\uparrow$ indicates the bigger the better while $\downarrow$ indicates the smaller the better. The best result is highlighted in bold
\end{table*}

\subsection{Ablation Study and Analysis}
\subsubsection{Effects of Different Components} 
\label{Effects of Different Components}
To better study the influence of each component in the proposed NORM-TR,  Table~\ref{tab: Effects of different Components} reports the ablation results of the subtraction of each component to the NORM-TR framework on the MOSI and MOSEI datasets, respectively. It is worth noting that the proposed NORM-TR has achieved state-of-the-art performance. 
As shown in the table, subtracting the NRGF/MF extractor decreases the accuracy to suboptimal performance, demonstrating the importance of NRGF and MF extractors for robust multimodal emotion recognition. 
After the removal of the multimodal fusion Transformer, the performance drops again, demonstrating that the multimodal fusion Transformer can effectively model the relations between NRGFs and MFs. 
Finally, the performance drops significantly when the noise-aware learning scheme is removed, especially for Acc-7. 
This suggests that the noise-aware learning scheme can help model to learn more useful emotion semantics from the multimodal data.
In addition, we also observed that the performance degradation is more significant when the NRGF extractor is removed than when the MFs extractor is removed. The possible reason for this is that MFs contain more noisy information.

\subsubsection{Effects of Different Modalities} 
To discuss the effect of each modality on performance, Table~\ref{tab: Effects of different Modalities} presents the ablation results of different modality settings on MOSI and MOSEI datasets, respectively. We observe that the combination of video, audio, and text information provided the best performance, suggesting that our model can learn the effective multimodal emotion representation for robust emotion recognition. On both datasets, the performance sharply dropped when the text modality was removed, indicating that the text modality plays an important role in multimodal emotion recognition. 

In addition, we tried to include different levels of Type-2 noise in each modality respectively on MOSI dataset, to discuss the sensitivity of each modality to noise. We found that the performance degradation was more significant when adding the noise to the text modality (\textit{e.g.,} decreased by 13\% on Acc-7 at 50\% mask percentage) than the audio and video modalities (\textit{e.g.,} decreased by 0.6\% and 1.1\% on Acc-7 separately at 50\% mask percentage). It shows that the text modality is more sensitive to noise than the other modalities.

\begin{table}[htpb]
	\begin{minipage}{0.48\textwidth}
		\small
		\caption{Effects of different Modalities}
		\label{tab: Effects of different Modalities}
		\setlength{\tabcolsep}{0.5mm}{
			\begin{tabular*}{\linewidth}{c@{\extracolsep{\fill}}cccc}
				\toprule
				\multirow{2}{*}{Method} & \multicolumn{2}{c}{MOSI} & \multicolumn{2}{c}{MOSEI} \\
				& Acc-7 ($\uparrow$) & MAE ($\downarrow$) & Acc-7 ($\uparrow$) & MAE ($\downarrow$) \\ 
				\midrule
				\textbf{NORM-TR} &\textbf{48.5}&\textbf{0.698} & \textbf{54.6} & \textbf{0.528}\\ 
				w/o Audio & 43.6 & 0.761 & 50.6 & 0.655 \\
				w/o Video & 43.3 & 0.765 & 51.8 & 0.592 \\
				w/o Text & 18.1 & 1.410 & 41.2 & 0.831 \\
				\bottomrule
			\end{tabular*}
		}
		Note: for each evaluation metric, $\uparrow$ indicates the bigger the better while $\downarrow$ indicates the smaller the better. The best result is highlighted in bold
	\end{minipage}
\end{table}

\begin{table}[htbp]
	\caption{Effect of different Query, Key, and Value Setting in Transformer \label{tab: Effects of different Query, Key, and Value Setting}}
	\setlength{\tabcolsep}{0.5mm}{
		\begin{tabular*}{\linewidth}{c@{\extracolsep{\fill}}ccccc}
			\toprule
			\multirow{2}{*}{Q} & \multirow{2}{*}{K \& V} & \multicolumn{2}{c}{MOSI} & \multicolumn{2}{c}{MOSEI} \\
			& & Acc-7 ($\uparrow$) & MAE ($\downarrow$) & Acc-7 ($\uparrow$) & MAE ($\downarrow$) \\ 
			\midrule
			MFs & NRGFs & 47.1 & 0.706 & 53.2 & 0.539\\
			\textbf{NRGFs} & \textbf{MFs} & \textbf{48.5} & \textbf{0.698} & \textbf{54.6} & \textbf{0.529}   \\
			\bottomrule
		\end{tabular*}
		\newline
		Note: for each evaluation metric, $\uparrow$ indicates the bigger the better while $\downarrow$ indicates the smaller the better. The best result is highlighted in bold
	}
\end{table}

\subsubsection{Different Query, Key, and Value Settings in Transformer}
Table~\ref{tab: Effects of different Query, Key, and Value Setting} presents the experimental results of different query, key, and value settings in Transformer on the MOSI and MOSEI datasets, respectively. We find that the best performance of the model was obtained when using the NRGFs as the query and the MFs as the key and value. This demonstrates that our NORM-TR has the ability to extract a more generic and useful query for the Transformer to help achieve better emotion recognition performance. 

\subsubsection{Effects of the Hyper-parameter Settings in Transformer}
\label{sec: Effects of the Hyper-parameter Settings in Transformer}
\begin{figure}[htbp]
	\centering
	\includegraphics[width=0.48\textwidth]{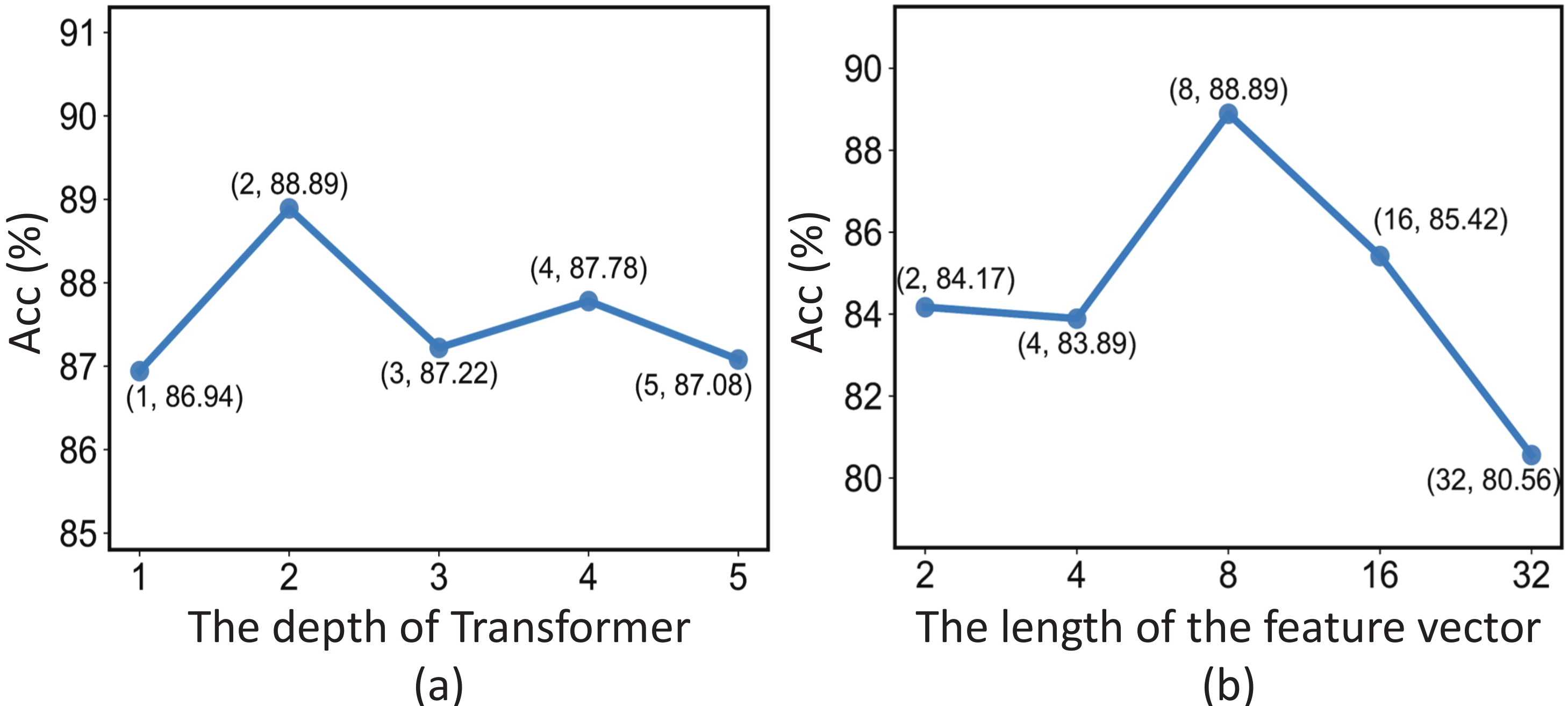}
	\caption{The impact of important parameter settings in Transformer on the RML dataset. (a) The effect of attention blocks in Transformer, and (b) the effect of the number of sampled frames in per sequence}
	\label{fig:Hyper_parameters_Selection}
\end{figure}

Fig.~\ref{fig:Hyper_parameters_Selection} (a) presents the accuracy of emotion recognition on the RML dataset, which is effected by the number of attention blocks in the Transformer architecture. As shown from the results, the accuracy achieved the highest 88.89\% when we set the depth to 2. Besides, we also observe that different Transformer depths only resulted in minor performance variations, indicating that the parameter has a small impact on our method.

In addition, we tried using more complex Transformer models instead of our multimodal fusion Transformer for fusion on the RML dataset, \textit{e.g.,} 1) using pairs of Transformers similar to MulT \citep{DBLP:conf/acl/TsaiBLKMS19} ; 2) concatenating the NRGFs and MFs and using a deeper ViT \citep{DBLP:conf/iclr/DosovitskiyB0WZ21} for fusion. We found that these complex models did not improve the performance, \textit{i.e.}, obtaining 88.75\% and 88.19\% of Acc-6 on the RML dataset by MulT and ViT, repectively. Meanwhile, these complex models require more parameters for training and additional computational cost (about 1 MACs). Our simple but effective multimodal fusion Transformer is able to complement the potentially insensitive but useful information of NRGFs by MFs containing more details, achieving more accurate emotion understanding (88.89\% of Acc-6).

\begin{figure*}[ht]
	\centering
	\includegraphics[width=0.7\textwidth]{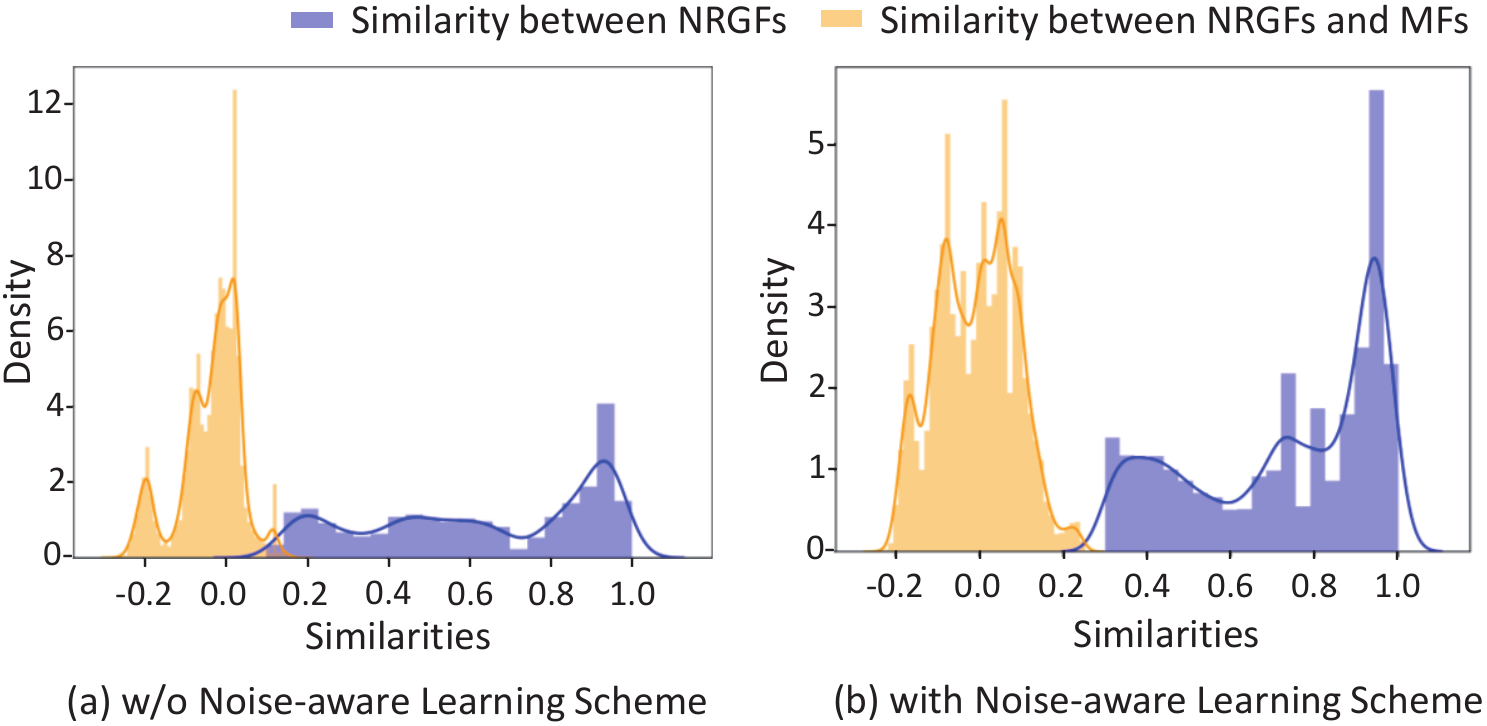}
	\caption{Visualization of NRGFs and MFs distributions on the MOSI dataset, with and without using the proposed noise-aware learning scheme, respectively. (a) Similarities distributions without using the noise-aware learning scheme on MOSI dataset; (b) Similarities distributions with using the noise-aware learning scheme on MOSI dataset}
	\label{fig:Modality_Representation}
\end{figure*}

\begin{figure*}[!ht]
	\centering
	\includegraphics[width=0.7\linewidth]{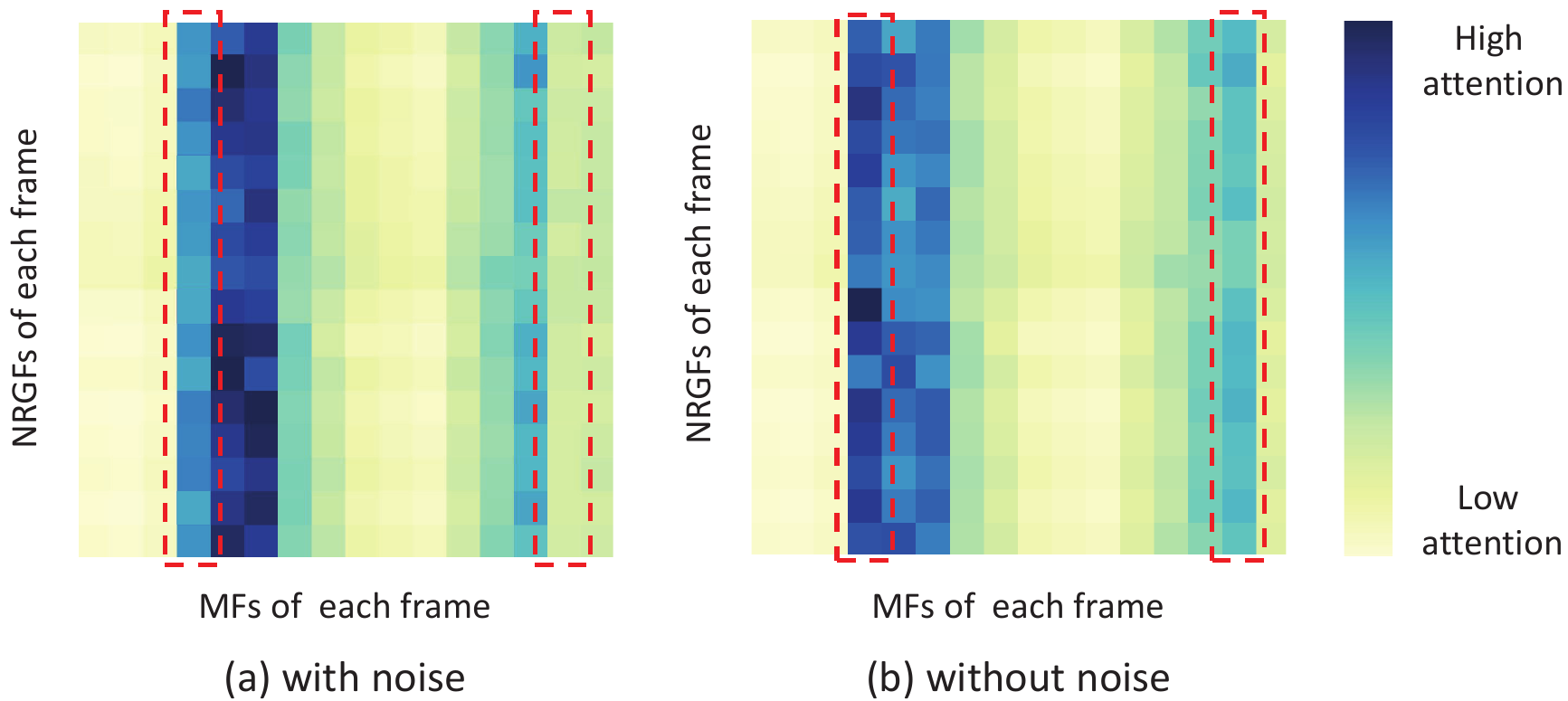}
	\caption{Visualization of the attention weights learned by the multimodal fusion Transformer for a randomly selected sample with and without a random mask on the RML dataset. (a) The attention weights with the mask noise, (b) the attention weights without the mask noise. Note: darker colors indicate higher attention weights for learning and the red dashed boxes represent the attention distribution of two randomly selected frames in the sequence}
	\label{fig:Attention_Weights}
\end{figure*}

\subsubsection{Effects of the Input Feature Vector Length}
In Fig~\ref{fig:Hyper_parameters_Selection} (b), we present the accuracy curves, which are effected by 
the length of the input feature vector, \textit{i.e.}, the number of frames sampled from an original sequence. As shown in the figure, the accuracy achieved the highest 88.89\% when we set the length to 8. Hence, we set the length of the input feature vector to 8. Moreover, the performance sharply dropped when the length was set to 32, indicating that too many frames sampled from data tend to introduce additional noise and lead to poor results. 

\subsubsection{Visualization of Noise-Resistant and Multimodal Feature Distributions}  
In Fig.~\ref{fig:Modality_Representation}, we visualized the similarity distributions of the extracted Noise-Resistant Generic Features (NRGFs) and Multimodal Features (MFs) on the MOSI dataset, with and without using the proposed noise-aware learning scheme, respectively. We applied the Kernel Density Estimation (KDE) and cosine similarity to describe the similarity distribution of the two types of features. As shown in Fig.~\ref{fig:Modality_Representation} (b), with a noise-aware learning scheme, most of the NRGFs are similar (similarities close to 1) while most of the NRGFs and MFs are different (similarities close to 0), demonstrating that the NRGFs are more generic and noise-consistent while MFs contain more modality-specific characteristics. On the contrary, without using the noise-aware learning scheme, the similarities between the NRGFs are significantly increased (see Fig.~\ref{fig:Modality_Representation} (a)), indicating that they retain more noisy information from different modalities.

\subsubsection{Visualization of Attention Weights Learned in Transformer}
Fig.~\ref{fig:Attention_Weights} (a) and (b) show the attention weight matrixes learned by the Transformer for a randomly selected sample with and without a random mask on the RML dataset, respectively. In practice, for audio and video data, we set the pre-computed feature vectors $U_a$, $U_v$ of the mask frame to 0. Since the NRGFs and MFs are shaped as $2T \times \frac{N}{2}$, where $T=8$, the size of the weight matrix is $16 \times 16$. Each square in the matrix represents the attention weight score learned by the Transformer between the corresponding frames of the NRGFs and MFs. The red dashed boxes show the attention distributions of two randomly selected frames in the matrix. Compared with the attention weights of the unmasked frames (obtaining lighter squares in Fig.~\ref{fig:Attention_Weights}(a)), the attention weights of the masked frames decrease significantly (obtaining darker squares in Fig.~\ref{fig:Attention_Weights}(b)). This indicates that the Transformer prefers to translate the information of the frame without the mask by extracting NRGFs as the query and MFs as the key and value, rather than focusing on noisy frames with the mask, so that suppress the side influence of noise information.

\section{Conclusion} \label{sec:Conc}
This paper proposed a novel Noise-Resistant Multimodal Transformer (NORM-TR) approach for multimodal emotion recognition. The NORM-TR consists of a Noise-Resistant Generic Feature (NRGF) extractor, a Multimodal Feature (MF) extractor, and a multimodal fusion Transformer to fuse NRGFs and MFs, thus significantly reducing the negative impacts of noise in the multimodal data. To this end, a novel noise-aware learning scheme is further designed to help optimize the NORM-TR appropriately to obtain noise-invariant emotion representations. Extensive experiments on several multimodal datasets, including MOSI, MOSEI, IEMOCAP, and RML, show that our method outperforms other approaches, demonstrating the importance of handling noisy information as well as the effectiveness of our method. 
Despite the effectiveness of our method, we found that our method does not capture the problem of the inconsistency and absence of emotional labels. In the future, we will introduce more advanced semi-supervised or self-supervised learning mechanisms into our method to learn from unlabeled data, thus obtaining a more robust emotion understanding.

{\small
\bibliographystyle{apacite}
\bibliography{sn-bibliography}
}

\end{document}